\documentclass[12pt]{elsarticle}
\usepackage{graphicx}
\usepackage{amssymb}
\usepackage[usenames]{color}
\def\msn{m_{\mbox{\tiny SN}}}
\def\msn{m_\nu}

\title{Neutrino mass bound in the standard scenario for supernova electronic antineutrino emission}
\author{G. Pagliaroli}
\address{INFN, Laboratori Nazionali del Gran Sasso, Assergi (AQ), Italy\\ICTP, Strada Costiera 11, I-34014 Trieste, Italy}
\author{F. Rossi-Torres}
\address{Instituto de F\'isica ``Gleb Wataghin'' - UNICAMP, 13083-970 Campinas SP, Brazil\\INFN, Laboratori Nazionali del Gran Sasso, Assergi (AQ), Italy}
\author{F. Vissani}
\address{INFN, Laboratori Nazionali del Gran Sasso, Assergi (AQ), Italy\\ICTP, Strada Costiera 11, I-34014 Trieste, Italy}

\date{LNGS/TH-01/10}                                           
\begin{document}

\begin{abstract}
Based on recent improvements of the supernova electron
antineutrino emission model, we update the limit on neutrino mass
from the SN1987A data collected by Kamiokande-II, IMB and Baksan.
We derive the limit of 5.8 eV at 95\% CL, that we show to be
remarkably insensitive to the astrophysical uncertainties. Also we
evaluate the ultimate mass sensitivity of this method for a
detector like Super-Kamiokande. We find that the bound lies in the
sub-eV region, 0.8 eV at 95 \% CL being a typical outcome,
competitive with the values that are presently probed in
laboratory. However, this bound is subject to strong statistical
fluctuations, correlated to the characteristics of the first few
events detected. We briefly comment on the prospects offered by
future detectors.

%

\end{abstract}
\maketitle

\centerline{\footnotesize \tt INFN preprint LNGS/TH-01/10}

\parskip7pt
%
\section{Introduction}
The interest in measuring the, presently unknown, absolute mass scale of neutrinos has been renewed
by the experimental evidences of neutrino oscillation \cite{conf,reviews}.

It is known since long \cite{z} that neutrinos from supernova can contribute valuable
information on the mass of neutrinos. In fact, the stringent limit of $m_\nu <5.7$ eV at
95 \% CL has been obtained by Loredo and Lamb \cite{ll} using SN1987A neutrinos \cite{k,i,b};
another important result in this connection is the one obtained by Nardi and Zuluaga, who
argue that future supernova will permit us to probe the sub-eV region \cite{nz1,nz2,nz3}.

In the present paper, we aim at updating both these results: namely, we improve
the bound on neutrinos from SN1987A and we evaluate the
ultimate sensitivity of the method to probe neutrino masses introduced by Zatsepin.

\section{The limit from SN1987A}
%
%

\subsection{The reasons of an updated analysis}
The limit from SN1987A \cite{ll} is quoted in the PDG report
but it is considered {\em ``no longer comparable with the limits from tritium
beta decay''} \cite{pdg}. In fact, in the 3 neutrino context  it
can be compared with the limit obtained in laboratory \cite{mainz,troitsk}; the value of the latter is $2$ eV, about three
times tighter than the former.
%
%

Nevertheless, the analysis on neutrino mass of Lamb and Loredo \cite{ll} maintains a big methodological merit, being the only one based
on a theoretically motivated model for the emission of neutrinos. Their model is capable of
reproducing the expected (main) features of neutrino emission and, in particular, it includes an initial
phase of intense luminosity. This phase, called accretion, is the crucial ingredient for theories that
attempt to explain the explosion of the star, based on the ``delayed scenario'' \cite{nad,bw}--see \cite{ds}
for a review. As we will show in the following, this phase is the theoretical ingredient that allowed to
obtain the comparably strong limit on the mass recalled previously.

There are two specific considerations that make a
reanalysis necessary: 1)~it has been noted that the likelihood function adopted by Lamb and Loredo has
a statistical bias \cite{liks}; 2)~in addition, an improved model for the emission of neutrinos (which
overcomes certain shortcomings and involves significant changes in the astrophysical parameters resulting
from SN1987A data analysis) has been recently introduced in the scientific literature \cite{ap,prl}.

\subsection{Procedure of analysis}
\bigskip
The method used in this paper to investigate the neutrino mass is
based on punctual comparison between the features of the collected
data \cite{k,i,b} and the expectations resulting from a specific theoretical
model~\cite{ap,prl}. This model describes the expected flux of
electron antineutrinos, taking into account that the main reaction leading
to observable events is $\bar\nu_e p\to e^+ n$ both in water
Cherenkov than in scintillator detectors.

We assume that the shape of the flux is known up to nine free
parameters that are obtained fitting the data. Let us explain their meaning:
The first six parameters belong to two emission phases (accretion and cooling) and are used to take into
account the large astrophysical uncertainties. Each emission phase is
characterized by its intensity, its duration and the average
energy of the emitted neutrinos.  The three parameters of
the accretion phase are the initial mass ($M_a$),
the time scale ($\tau_a$) and the initial temperature ($T_a$);
those of the cooling phase are the radius ($R_c$), the time scale ($\tau_c$) and the initial temperature ($T_c$).
For details and analytical expressions see \cite{ap,prl}.
The other three parameters are
called ``offset times''; each one of them is the absolute delay of
the first observed event in each detector, more explicitly described
below. We need to include three different offset times because the
clocks of Kamiokande-II, IMB and Baksan were not synchronized
\cite{k,b}.

\medskip
Now we include the effects of neutrino mass. The antineutrino
flux, $\Phi_{\bar\nu_e}(t,E_\nu)$, is a parametric function that
depends on the time of emission ($t$) and on the energy of the
antineutrino ($E_\nu$), see in particular Eqs.~10, 13, 19 and 20
in reference \cite{ap}. Of course this function must vanish for
$t\le 0$.
Using the same notation of ref.~\cite{ap} (see in particular Eq.~8 there)
we can write the emission time for the $i$-th event as follow:
\begin{equation}
t_i=\delta t_i + t^{\mbox{\tiny off}}-\Delta t_i.
\label{1}
\end{equation}
The first term in the right hand side, $\delta t_i$, is the
relative time between the $i$-th and the first observed event in
the considered detector, which is known directly from the data
without significant error. The second one, $t^{\mbox{\tiny off}}$, is the offset time
parameter which is the sum of the emission time of the first neutrino
detected, $t_1$, and of its delay due to the velocity of
propagation, $\Delta t_1$, namely $t^{\mbox{\tiny off}}=
t_1+\Delta t_1$. Finally, the last term,
\begin{equation}
\Delta t_i=\frac{D}{2c}\left(\frac{\msn}{E_{\nu, i}}\right)^2,
\label{del}
\end{equation}
is the delay of the neutrino due to a non-zero mass \cite{z}, where
$D$ is the distance of propagation.
The neutrino energy $E_{\nu, i}$ of the $i$-th event can be 
reconstructed from the measured energy of the positron, $E_i$,
which is known up to its error, $\delta E_i$.
The numerical value of the delay
when ${D}={\mbox{50 kpc}}$ (as for SN1987A),
$E_\nu={\mbox{10 MeV}}$ (a typical value)  and
$m_\nu={\mbox{10 eV}}$  is $\Delta t=2.6$ s,
which is five times longer than the duration
of the phase of accretion.


The scope of the statistical analysis is to extract from the fit
$t^{\mbox{\tiny off}}$ and ${\it \msn}$ at the same time. It is quite
evident that these two terms work in opposite sense,
see Eq.~(\ref{1}) and recall the condition $t_i \ge 0$.
This makes the extraction of these two parameters more difficult, especially in the case of
SN1987A, when the number of observed events is small.
We adopt the same likelihood function $\mathcal{ L}$ constructed
in \cite{ap} including in it the expression for the
times $t_i$ given in Eq.~(\ref{1}). This is a function of 10 parameters, namely the
nine parameters previously discussed plus the neutrino mass.
\begin{figure}[tb] 
   \centering
   \includegraphics[width=7.5cm]{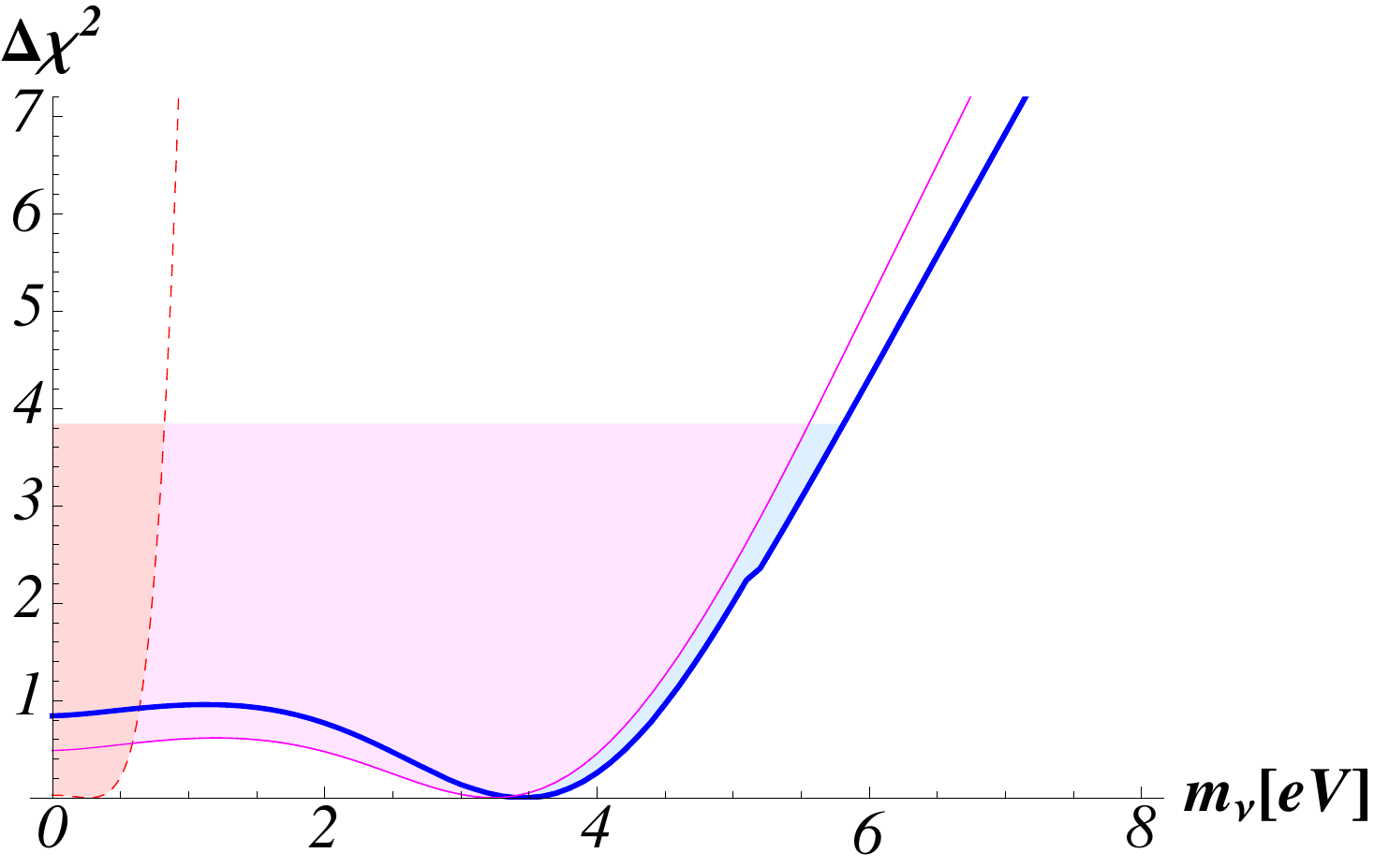}
   \caption{\em\small The curves show various $\Delta\chi^2(\msn)$ obtained by analyzing
   supernova neutrino data as a
   function of the neutrino mass. The two continuous curves are obtained from
   SN1987A data; the thick one includes the astrophysical uncertainties,
   the thin one assumes instead that the astrophysical parameters of neutrino emission are known.
   For comparison, we include the result of the analysis of simulated data set, collected in a detector
   a la Super-Kamiokande (SK), for a future supernova exploding at 10 kpc from us (leftmost dashed curve).
   This curve, discussed in detail later, illustrates the ultimate sensitivity of the method.
   }
   \label{f}
\end{figure}

\subsection{Results and remarks}
Using the definition $\mathcal{L}=\exp(-\chi^2/2)$
we obtain the function that allows us to estimate the neutrino mass
\begin{equation}
\Delta\chi^2(\msn)=\chi^2(\msn)-\chi^2_{\mbox{\tiny best fit}},
\end{equation}
This function is plotted in Fig.~\ref{f}. The two continuous
curves show the results of SN1987A data analysis. The thick line
is obtained, for any fixed value of the neutrino mass, maximizing
the likelihood with respect to the other 9 free parameters. The
curve is somewhat bumpy, reflecting the presence of multiple
maxima that compete in the likelihood with similar height. The
existence of these maxima has been already remarked in \cite{ap}
and causes numerical difficulties. To avoid these problems we
bound the mass of accreting material $M_a$, which regulates the
intensity of neutrinos emission in the accretion phase, to be
lower than 0.6 $M_\odot$ \cite{ll,ap}. The thin line, instead,
arises when the 6 astrophysical parameters are set to the best-fit
values obtained in \cite{ap}. In this case only the 3 offset times
are allowed to fluctuate freely to maximize the likelihood. The
comparison of the two curves reveals some interesting features:
\begin{itemize}
\item In spite of the really different assumptions, namely the
complete knowledge of the supernova $\bar\nu_e$ emitted flux or
only of its shape (up to 6 parameters), the two curves are quite
similar. This shows that the large uncertainties in the
astrophysics of the emission {\em are not} the main limitation in
this type of analysis.
\item In both curves, the minimum is
located at $\msn\neq 0$, however, this is not statistically
significant.\footnote{We note that the
presentation using $\msn$ rather than $\msn^2$--which is the
quantity that is actually probed, see Eq.~(\ref{1})--emphasizes
somewhat artificially the region close to $\msn\sim 0$.}
This is linked to a clustering of the events \#1,2,4,6 of
Kamiokande-II for $\msn\sim 3.5$ eV, already remarked
by several authors, e.g., \cite{huzita}.
\end{itemize}

{}From our statistical analysis, we obtain as limit on neutrino mass from SN1987A data
the value
\begin{equation}
\msn<5.8\mbox{ eV at 95\% CL}.\label{bondi}
\end{equation}
As already noted, this does not change much if we assume that the
astrophysics of the emission is perfectly known; in this case, in
fact, the limit becomes $\msn<5.6\mbox{ eV at 95\% CL}$: see
Fig.~\ref{f}. 

{}From Eqs.~(\ref{1}) and (\ref{del}) and from the previous discussion,
it is quite evident that the information on the
presence of the neutrino mass is
mainly contained in the earlier events and, in particular, in
those with low energy. These considerations select, as the most
relevant data, the six events collected by Kamiokande-II in the
first second~\cite{k}, that incidentally, are also the most
relevant ones to determine the presence of an
accretion phase~\cite{ll,ap}.

\section{The sensitivity of the method}

These findings led us to the question of evaluating the ultimate
sensitivity of this method for a future galactic supernova event.
For this aim, we will analyze in this section simulated data,
extracted from the generator described in \cite{prl} upgraded to
describe the propagation of massive neutrinos, focussing mostly on
the possibilities of the  existing detectors. We will introduce
and critically examine the assumptions used to derive the bound,
comment on their statistical meaning, and overview the prospects
offered by future detectors.

\subsection{Statistical procedure}
The expected counting rate of the signal is a function
of the emission time $t$, neutrino energy $E_\nu$, detector mass $M_d$,
distance of the supernova $D$ and of the astrophysical parameters that describe the
electron antineutrino emission,
namely
\begin{equation}
R(t,E_\nu)=6.7\times 10^{31}
\frac{M_d}{1 \mbox{ kton}}
\sigma_{\bar\nu_e
p}(E_\nu)\tilde\Phi_{\bar\nu_e}(t,E_\nu)\epsilon(E_{e^+}),
\label{rate}
\end{equation}
that depends on the supernova distance through the electron
antineutrino flux, i.e., $\Phi_{\bar\nu_e}\propto 1/D^2$. We will
consider a SN exploding at a distance of $D=10$ kpc, typical of a
galactic event \cite{vis,mir}. Here, $\sigma_{\bar\nu_e p}(E_\nu)$
is the cross section of the interaction process~\cite{visstru};
the function $\epsilon(E_{e^+})$ is the detector efficiency that
we set to $98\%$ above a threshold of 6.5 MeV; we approximate
$E_e=E_\nu- \Delta$ with $\Delta=1.293$~MeV.

Finally, $\tilde\Phi_{\bar\nu_e}(t,E_\nu)$ in Eq.~(\ref{rate})
is the same electron antineutrinos flux used previously, $\Phi_{\bar\nu_e}(t,E_\nu)$,
improved taking into account the finite rising time of the signal.
This is described by an exponential function characterized by a new time scale,
$\tau_r > 30$~ms, that we treat as a new parameter of the analysis~\cite{prl}.
With future large statistics we will be able to probe such a small
time structure, as argued in \cite{prl}. Thus, in our analysis
this function depends on 7 astrophysical parameters.

\begin{figure}[tb] 
   \centering
   \includegraphics[width=0.45\textwidth]{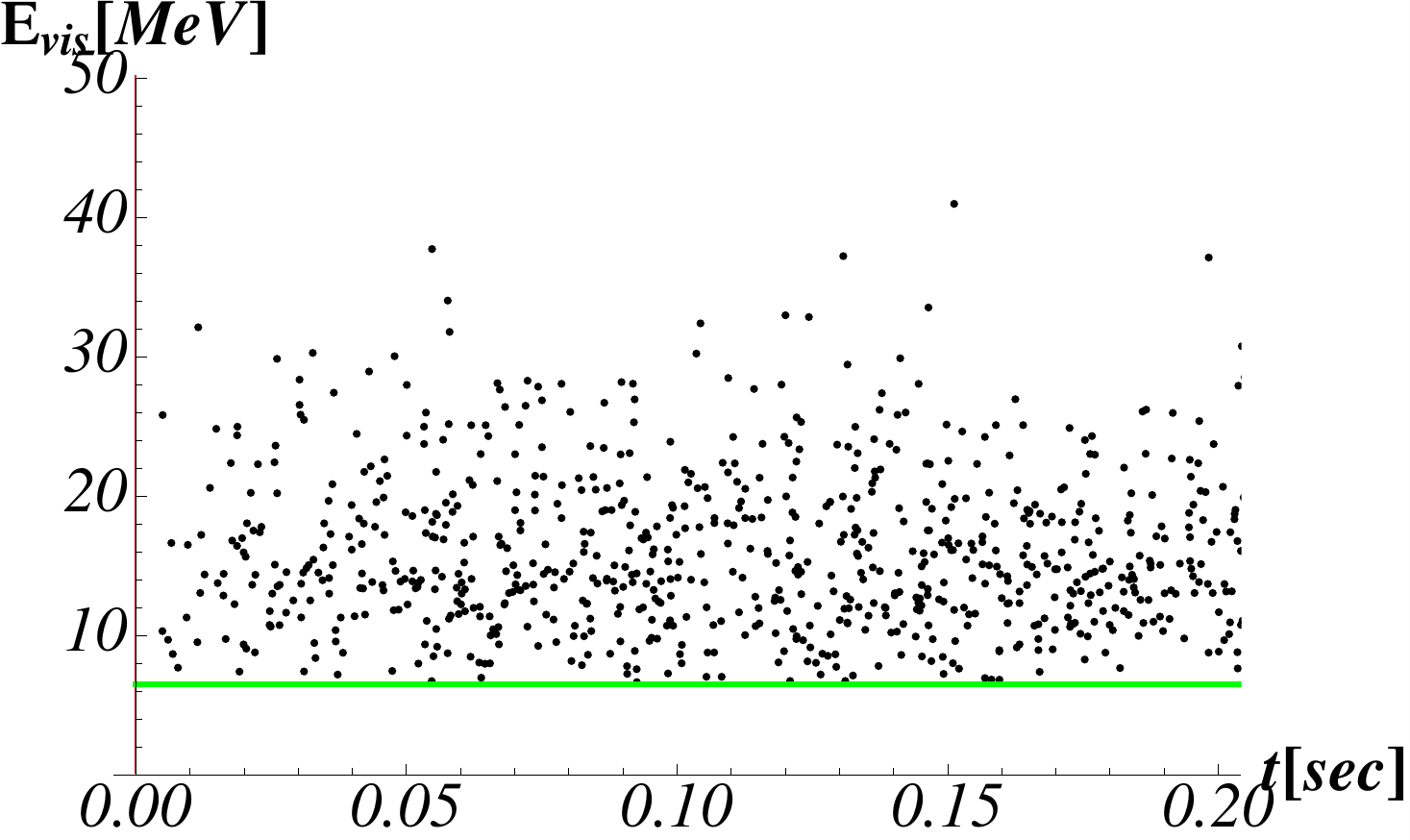}\includegraphics[width=0.45\textwidth]{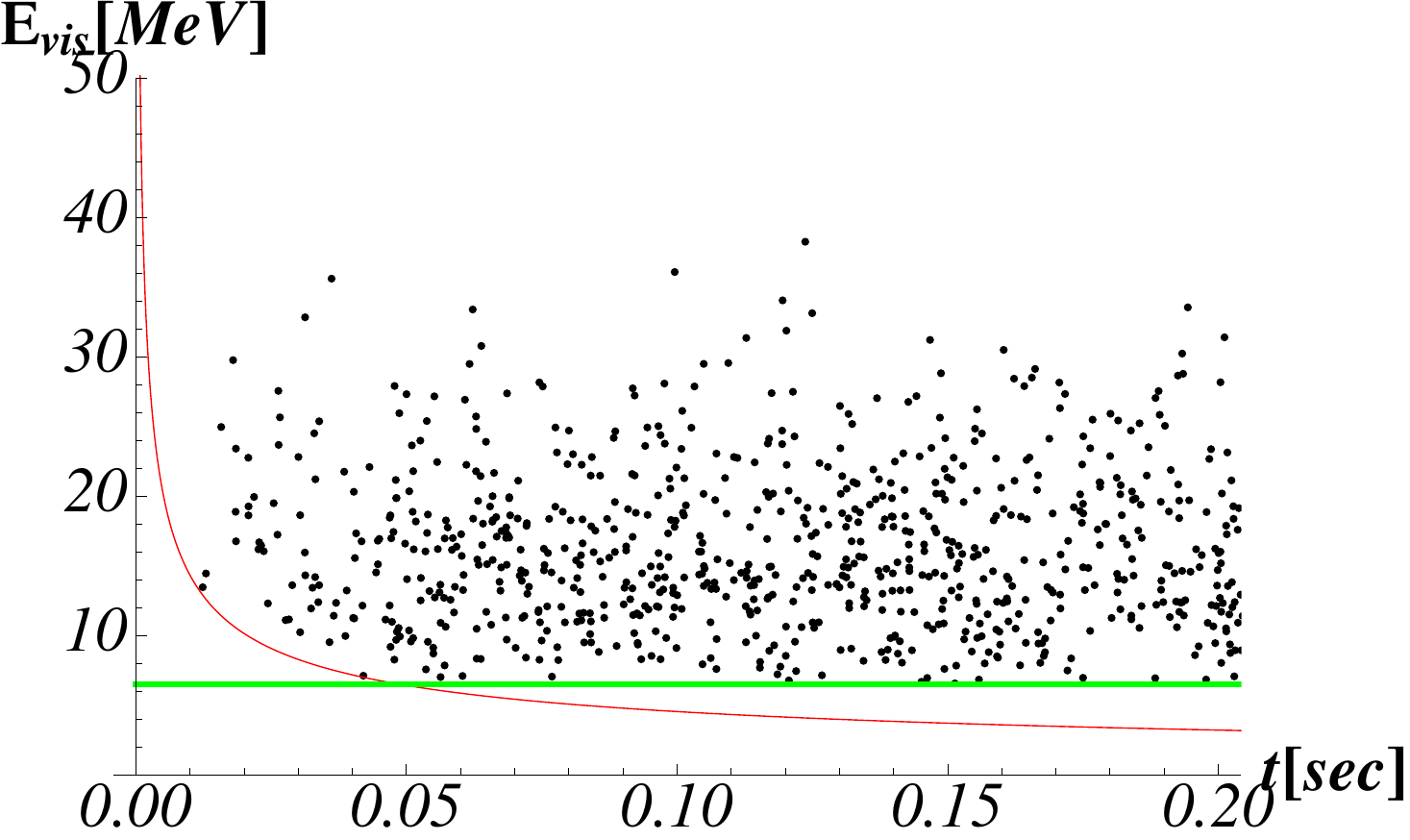}
   \caption{\em\small Positron energy versus the emission time for two samples of
   simulated events for Super-Kamiokande detector.  The neutrino mass is set to zero
   in the left panel and to the bound from tritium decay, $m_\nu=2$ eV in the right one.
   The green line is the threshold of the detector and the red curve
   is the expected delay due to the neutrino mass, Eq.(\ref{del}).}
   \label{gen}
\end{figure}

Each event extracted from this function is characterized by its
relative detection time $\delta t_i$, its positron energy, $E_i$,
and the error on this energy given by the function $\delta
E_i/E_i=0.023+0.41 \sqrt{\mbox{MeV}/E_i}$ \cite{sk}. We generate
the data using the Monte Carlo described in \cite{prl} and take
into account the effect of neutrino mass by assigning a time delay
to each generated event, as prescribed by Eq.~(\ref{del}).
Fig.~\ref{gen} shows two extractions, magnified in the region of
the first 200 ms of data taking. Their comparison shows clearly
the region where the effect of neutrino mass is most relevant,
namely the one with the lowest energies and the smaller detection
times.

We studied ten simulated data sets  for a detector with a fiducial mass of $M_{SK}=22.5$ kton
as the Super-Kamiokande detector, which corresponds to about $4482$ events on average.
We also considered two different detector masses: $M_d=M_{SK}/16$, with an average number
of events of $280$, similar to the ones expected on LVD detector, and also $M_d=M_{SK}/256$,
corresponding to an average number of $18$ expected events, which resembles the statistics collected for SN1987A.

We calculate the bound on the mass by assuming that:
\begin{itemize}
\item The astrophysical parameters are precisely known; we use
those in ref.~\cite{ap}, that agree with the expectations of a
standard collapse and set $\tau_r=50$~ms. \item The offset
time is known without significant error (more discussion later).
\end{itemize}
These are very optimistic assumptions as appropriate to evaluate
the ultimate sensitivity of the method. We discuss the weight of
these assumptions in the following and their implications on the
understanding of SN1987A results.

We note in passing that the neutrino mass enters the likelihood
through Eq.~(\ref{del}), in the form $\msn^2 D$; moreover, the
interaction rate in Eq.~(\ref{rate}) depends on the combination
$M_d/D^2$: thus, the likelihood obeys the exact scaling law
\begin{equation}
\mathcal{ L}(M_d,D,\msn)=\mathcal{ L}(\alpha^2 M_d,\alpha D,\msn
/\sqrt{\alpha}). \label{scalon}
\end{equation}
This means, e.g., that once we know the value of the neutrino mass
bound for  $M_d=M_{SK}/16$, we get the bound for $M_d=M_{SK}$ and
$D=40$~kpc simply halving~it. {}From here, we also conclude that
the bound on the mass can be written as
$m_\nu<f(M_d/D^2)/\sqrt[4]{M_d}$, where the function $f$ depends
on the selected statistical level on the adopted test and on the
specific data set.

\begin{figure}[tb] 
   \centering
   \includegraphics[width=7.5cm]{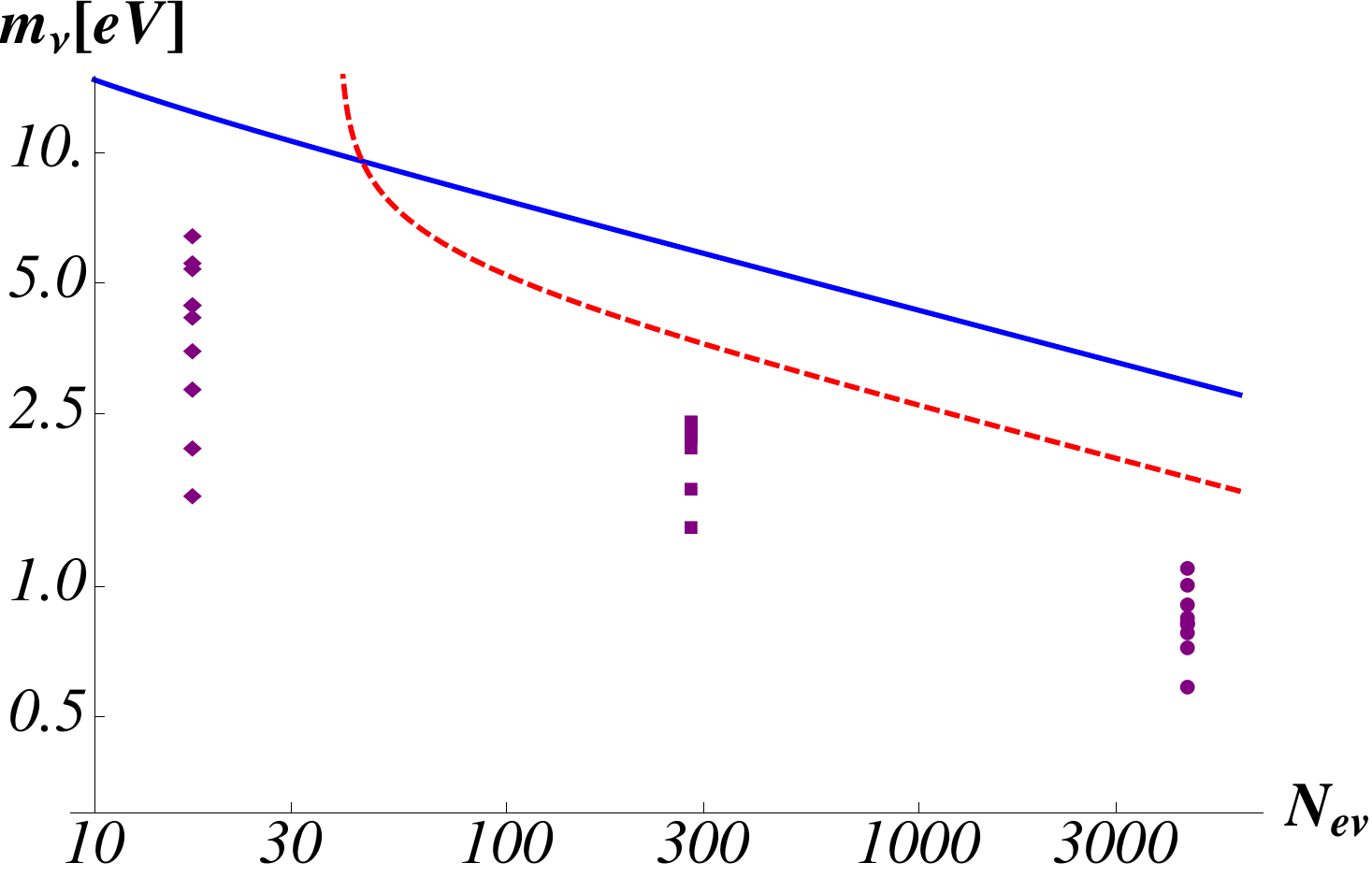}
   \caption{\em\small The dots represent the 95\% CL bounds
   on neutrino mass from the analysis of simulated data; for
   each of the three value of the average numbers of expected events we extracted
   and analyzed 10 simulated data set. Circles (dots in the right), squares (center) and
   diamonds (left) correspond to the results in detectors with masses
   $M_d=M_{SK}$, $M_{SK}/16$ and $M_{SK}/256$, respectively.
   The continuous and dashed curves 
   describe the bounds given by Eq.~(\ref{eq8}) and discussed in the text.}
   \label{f2}
\end{figure}

\subsection{Results and discussion}
The $95\%$ CL neutrino mass bounds, obtained with fixed
astrophysical parameters, are reported in Fig. \ref{f2} for each
analyzed data set.

\paragraph{The case of low statistics and SN1987A}
As first step, we discuss the diamonds points corresponding to a
detector with mass $M_d=M_{SK}/256$. The average number of events
in this case is very similar to the one observed for SN1987A, so
we can explore the fluctuations due to the features of the data in
a small data set. The values of the neutrino mass bound fall in
the interval of $1.6 \mbox{ eV}<\msn<6.4 \mbox{ eV}$ showing that
each particular data set contains very different information about
the neutrino mass presence.
%

We used these simulations to investigate the weight of the various
assumptions of the analysis on the resulting bound. For a typical
simulated data set, we analyzed the data using three different
procedures:

\begin{enumerate}

\item We suppose to know all the astrophysical parameters without
errors and also the offset time. In other words, the only free
parameter of the likelihood is the neutrino mass. The resulting
95\% CL on neutrino mass in this case is: $\msn <4.4$ eV.

\item We suppose that the offset time is unknown, whereas the 7
astrophysical parameters are known from the theory. Namely, the
likelihood is a function of the neutrino mass and of the offset
time. The resulting 95\% CL on neutrino mass in this case is:
$\msn <7.2$ eV.

\item Finally, we suppose that we do not know any of the 9
parameters. Namely, only the shape of the signal in known from the
theory and all parameters have to be estimated from the likelihood
analysis. The resulting 95\% CL on neutrino mass in this case is:
$\msn <7.4$ eV.
\end{enumerate}

This study shows that the knowledge of the offset time significantly affects the value of the mass bound.
Instead the comparison of the last two results confirms that the knowledge of the astrophysical parameters
is less critical for the analysis, in agreement with what we found for SN1987A data analysis.

We are ready for the comparison with the SN1987A results. This is
possible using the scaling relation of Eq.~(\ref{scalon}). Using
$\alpha=5$, we translate the range $1.6-6.4$ eV in the range
$0.7\mbox{ eV}<\msn<2.9$ eV when $D=50$ kpc and for a detector
mass $M_d\simeq  M_{KII}$. For a fair comparison, we still need to
take into account that the offset times and the astrophysical
parameters are unknown in SN1987A analysis. So, comparing $\msn
<4.4$ eV and $\msn <7.4$ eV, we multiply this range by the factor
$7.4/4.4$ obtaining $1.2\mbox{ eV}<\msn<4.9$~eV. The bound from
SN1987A, $\msn<5.8$ eV, is not far from this range. The residual
difference can be attributed to the better performances of the
simulated detector. In fact, the improved efficiency implies that
more events are collected at low energies; moreover, any
misidentification of events is forbidden by constructions, due to the
postulated absence of background events above the detection threshold.

\bigskip

\paragraph{The case of high statistics and the ultimate upper limit}
Now we discuss the results for high statistic, i.e., the case when
$M_d=M_{SK}$, chosen to represent the observation of a future
galactic supernova event. A typical simulated data set is the one
shown in the left panel of Fig.~\ref{gen}; the resulting $\Delta
\chi^2$ is the one given by the dashed line of Fig.~\ref{f}, that
implies:
\begin{equation}
\msn<0.8\mbox{ eV at 95\% CL}
\label{bondo}
\end{equation}
This result confirms the possibility to probe the sub-eV region
for neutrino mass using SNe neutrinos, in agreement with the
finding of Nardi and Zuluaga \cite{nz1,nz2,nz3}. Also it would be
closer to the sensitivity of about $0.2$ eV that will be probed by the
Katrin experiment \cite{kat}.
\bigskip

%
%

An inspection of Fig.~\ref{f2} shows also quite clearly that the
bounds fluctuate strongly with the individual simulation, even for
high statistics. This can be explained as follows. The emission
time of each signal event is subject to the condition $t_i>0$ (see
Eq.~(\ref{1})), which implies the condition on the neutrino mass:
\begin{equation}
\label{sa} m_\nu< m_\nu^* =\mbox{Min}_i \left\{ E_{\nu,i}
\sqrt{\frac{t^{\mbox{\tiny off}}+\delta t_i}{D/(2 c)}} \right\}.
\end{equation} When we replace $E_{\nu,i}=E_i+\Delta$, namely,
neglecting the error in the measurement of the positron energy, we
obtain the bound $m_\nu^*$ on the neutrino mass directly from the
data. Typically, the minimum in Eq.~(\ref{sa}) corresponds to the
first (or the first few) event(s) of the data set; compare, e.g.,
with Fig.~\ref{gen}. This means that the role of the fluctuations
is very important, also for large number of detected events. In
other words, the bound $m_\nu^*$ depends strongly on the specific
data set. We compare this bound with the one obtained by the full
analysis of the likelihood function in Tab.~\ref{tab3}. Within
25\% the two bounds are in agreement. This supports the idea that,
in this type of analysis where $t^{\mbox{\tiny off}}$ is known,
the information on the neutrino mass is mostly contained in the
first few events, rather than somewhat distributed in the data
set.

\begin{table}[t]
\begin{center}
\begin{tabular}{|c||c|c|c|c|}
\hline
$N_{ev}$ & $t^{\mbox{\tiny off}}$(ms) & $\msn $(eV) & $\msn^*$(eV)
\\
\hline\hline
$4328$ & $4.9$ & $0.78$ & $0.94$  \\
\hline
$4479$ & $2.9$ & $0.82$ & $0.68$   \\
\hline
$4497$ & $2.9$ & $0.72$ & $0.76$   \\
\hline
$4473$ & $2.9$ & $1.10$ & $1.10$  \\
\hline
$4492$ & $4.6$ & $0.84$ & $0.77$  \\
\hline
$4464$ & $2.9$ & $0.82$ & $0.73$  \\
\hline
$4488$ & $3.5$ & $0.90$ & $0.72$  \\
\hline
$4412$ & $2.6$ & $0.82$ & $0.65$  \\
\hline
$4412$ & $1.2$ & $0.59$ & $0.52$  \\
\hline
$4399$ & $3.0$ & $1.01$ & $0.77$  \\
\hline
\end{tabular}
\caption{\em\small Number of events in Super-Kamiokande, offset time,
statistical bound on the neutrino mass, and neutrino mass bound from Eq.~(\ref{sa}) in 10 simulations.}\label{tab3}
\end{center}
\end{table}

\paragraph{An alternative estimator}
For comparison, we present also other estimators of neutrino
masses that, instead, depend on relatively large numbers of
events. We construct them by imposing that the error on a typical
time scale of neutrino emission, ~$\tau$, is larger than the
average delay of the events $\langle{\Delta t}\rangle$:
\begin{equation}
\label{eq7}
\langle{\Delta t}\rangle < n_\sigma \cdot \frac{\tau}{\sqrt{N-1}},
\end{equation}
where $n_\sigma$ is the required sensitivity (the number of sigmas);
$N$ is the number of detected events in the assigned time
scale $\tau$ ($N\gg 1$ since we want to determine experimentally the phase of emission);
$\langle{\Delta t}\rangle$, in turn, will be derived from Eq.~(\ref{del})
replacing the neutrino energy with its average value $\langle E_\nu \rangle$.
For a similar proposal, see \cite{beac}. Setting
$n_\sigma=2$, we get  the following bound on the neutrino mass:
\begin{equation}
m_\nu <\frac{2\langle E_\nu
\rangle}{\sqrt[4]{N-1}}\sqrt{\frac{\tau}{D/c}}
\label{eq8}
\end{equation}
We use the value $\langle E_\nu \rangle=13$ MeV and $D=10$ kpc for
numerical purposes and consider two concrete times scales of
emission: the one of the accretion phase, $\tau=\tau_a=0.55$~s,
which corresponds to $N=0.4 N_{ev}$; the one of the rising
function, $\tau=\tau_r = 50$~ms, which corresponds instead to
$N=N_{ev}/40$. These lead to the continuous curve  and to the
dashed curve of Fig.~\ref{f2}, respectively. As soon the expected
number of events is large enough ($N\gg 1$), we get a stabler
bound on the neutrino mass.\footnote{Note that in this limit and
considering that the number of events scales as $1/D^2$, the bound
in Eq.~(\ref{eq8}) is {\em independent} from the distance. The
same occurs with the bound in Eq.~(\ref{sa}) if $t^{\mbox{\tiny
off}}+\delta t_i\propto D/\sqrt{M_d}$, that is satisfied for an
initial linear rise of the interaction rate, $R(t)\simeq \xi t
M_d/D^2$ (which is the quantity that determines the time of the events).} However, Fig.~\ref{f2} shows clearly that these are
very conservative upper bounds, when compared with the true bounds
from the full likelihood analysis.

\medskip

\paragraph{The gravity wave trigger and its limitations}
An important remark is in order. We evaluated the ultimate
sensitivity of the method with the existing neutrino detectors,
assuming that the offset time was known without significant error.
How can we achieve this? In principle, we could profit of the
detection of gravity waves, assuming they will be seen. 
However, two additional conditions
should be fulfilled: a precise location of the supernova in the
sky is needed; the interval of time between the onsets of
gravitational and neutrino emissions should be known. The first
condition is needed if the detectors of gravity waves and of
neutrinos are not in the same location. Elastic scattering
neutrino events can provide such an information, but with a
uncertainty of several ms \cite{prl}, while an astronomical
identification would make this error negligible. The second
condition has at present an associated theoretical $1\sigma$ error
of about 1~ms \cite{prl}, which is already limiting the
sensitivity of the existing neutrino detectors: see
Tab.~\ref{tab3}, or consider that IceCUBE uses 2~ms time window.
In summary, the key condition for a successful search for neutrino
mass by this method is the possibility to implement very precise
measurements  of the time; however, the previous discussion showed
the difficulties to realistically surpass the millisecond time
scale. It will be important to take into account these
considerations for an analysis of future real data, for instance,
taking into account the errors on the measurement of time. Another
way to go beyond these limitations would be to rely on larger
samples of data; this will be possible by future detectors, which
leads us to the last point of the discussion.

\medskip

\paragraph{Future prospects}
Finally, we comment on the prospects to improve the reach of this
method to investigate neutrino masses. A straightforward
possibility would be to use a bigger detector, say of megaton
mass; note incidentally that this is mentioned already in the
paper of Zapsepin \cite{z}. For example, with an increase of the
number of events expected in Super-Kamiokande ($22.5$ kton
fiducial volume) by a factor of $\sim 20$ one could expect an
improvement on $\msn^2$ as the inverse of the square root of this
number, thus reaching $\msn \le 0.4$ eV in the most optimistic
case. If instead we use the stabler bound of Eq.~(\ref{eq8}), we
find again a value close to the one in Eq.~(\ref{bondo}).

An alternative possibility would be to identify the very short
burst from early neutronization; see \cite{arnaud} for an earlier
discussion. Its detection could permit us to investigate neutrino
masses of similar size. In the standard scenario of neutrino
emission, however, this burst leads to a very small fraction of
the total number of events, which leads us again to consider a
megaton water Cherenkov detector. In fact, the elastic scattering
events are 1/35 of the total sample; the burst comprises some 1/20
of the total energy released in $\nu_e$'s, which, when converted
in $\nu_{\mu,\tau}$'s by oscillations, have a cross section 6.5
times smaller. Thus, a conservative estimation of the event
fraction from the neutronization burst is 1/4500, which means
about $N=$20 (=1) events in 450 (22.5) kton of fiducial volume
from a supernova at nominal distance of 10~kpc. If used in
Eq.~(\ref{eq8}) with $\tau=3$ ms, this yields $m_\nu\le 0.7$ eV,
which is again similar to the bound of Eq.~(\ref{eq8}), but
possibly more stable and without resorting to the gravity wave
trigger.\footnote{For a more precise bound one should keep into
account that the elastic scattering reaction $\nu e \to \nu e$
does not allow to reconstruct the neutrino energy precisely.}

\section{Summary}
The present work, part of a series of papers on supernova
neutrinos \cite{liks,ap,prl,vis,visstru}, was devoted to derive and discuss the bound
on neutrino mass from supernova electron antineutrinos. Our bound,
Eq.~(\ref{bondi}), agrees well with the one obtained by Lamb and
Loredo \cite{ll}, despite the large number of differences in the
procedures of analysis.

We argued that the result from SN1987A is relatively insensitive
to the details of the emission model, as soon as the emission
resembles the expectations of the standard scenario, that includes
an initial phase of intense antineutrino luminosity. We showed
that the knowledge of the time when neutrino emission begins
(`offset time') has, instead, a significant impact on the bound
that the existing detectors can obtain.


We derived the ultimate sensitivity that can be provided by
supernova neutrinos with existing detectors.  We showed that on
average it lies in the interesting sub-eV region. However the key
role of the first few detected events also implies a large
fluctuation on the mass bounds. A crucial requirement is the need
to reach very precise measurements of the offset time; we argued
that the detection of a gravity wave burst could permit to reach
the sub-eV sensitivity with the existing neutrino detectors. We
briefly commented on the prospect to improve the bound using
future, megaton class, water Cherenkov detectors.



\subsection*{Acknowledgments}
F.~Rossi Torres thanks CAPES and FAPESP for financial support. 
The work of G.~Pagliaroli is supported by 
``Fondo F.S.E.\  del Piano Operativo 2007-2008 del POR Abruzzo 2007-2013''.

\end{document}